\documentclass[prd,showpacs,floatfix,amsmath,nofootinbib,amssymb,floatfix,twocolumn]{revtex4}
\usepackage{graphicx,color,dcolumn,booktabs,bm,multirow}
\usepackage{longtable,lscape,footnote}
\usepackage{txfonts}
\usepackage{overpic}
\usepackage{amssymb}
\usepackage{array}
\usepackage{subfigure}
\usepackage{indentfirst}
\usepackage{enumitem}
\usepackage{feynmf}   
\usepackage{slashed}  
\usepackage{cases}
\usepackage{color}
\usepackage{epstopdf}
\usepackage[colorlinks,
            citecolor=green,
            anchorcolor=red,
            menucolor=red,
            linkcolor=red,
            filecolor=red,
            runcolor=red,
            urlcolor=blue,
            frenchlinks=red]{hyperref}
\usepackage{ulem}
\usepackage{tikz}
\usepackage{lipsum} 

\begin{document}
\title{General mathematical analysis on multiple solutions of interfering resonances combinations}
\author{Yu Bai}\email{baiy@seu.edu.cn}
\author{Dian-Yong Chen\footnote{Corresponding author}} \email{chendy@seu.edu.cn}
\affiliation{School of Physics, Southeast University, Nanjing 210094, China}
\newcommand{\tabincell}[2]{\begin{tabular}{@{}#1@{}}#2\end{tabular}} 
\def\TeV{\ifmmode {\mathrm{\ Te\kern -0.1em V}}\else
                   \textrm{Te\kern -0.1em V}\fi}%
\def\GeV{\ifmmode {\mathrm{\ Ge\kern -0.1em V}}\else
                   \textrm{Ge\kern -0.1em V}\fi}%
\def\MeV{\ifmmode {\mathrm{\ Me\kern -0.1em V}}\else
                   \textrm{Me\kern -0.1em V}\fi}%
\def\keV{\ifmmode {\mathrm{\ ke\kern -0.1em V}}\else
                   \textrm{ke\kern -0.1em V}\fi}%
\def\eV{\ifmmode  {\mathrm{\ e\kern -0.1em V}}\else
                   \textrm{e\kern -0.1em V}\fi}%
\begin{abstract}
When fitting cross sections with several resonances or interfering background and resonances, one usually obtains multiple solutions of parameters with equal fitting quality. In the present work, we find the source of multiple solutions for a combination of several resonances or interfering background and resonances by analyzing the mathematical structure of the Breit-Wigner function. We find that there are $2^n$ fitting solutions with equal quality for $n+1$ resonances, and the multiplicity of the interfering background and resonances depends on zeros of the amplitudes in the complex plane. We provide a simple, general method to infer all other solutions with equal fitting quality from a known solution.

\end{abstract}

\date{\today}
\maketitle

\section{Introduction}

To date, most low-lying hadron states have been well established, and physicists have begun to examine highly excited states. Compared to low-lying states, mass gaps of excited hadrons are rather small, while the widths of these states are much larger. There are moments when the widths of the involved states are comparable to their mass gaps, thus interference between them can change the Breit-Wigner distribution of resonances. By contrast, resonance parameters obtained from the mass spectrum or cross sections should also depend on interference between the involved states. When one fits the experimental data with more than one resonance, one can usually obtain several different solutions with equally good quality, i.e., the $\chi^2$ or likelihood of different solutions are the same (e.g., see Refs. \cite{Chen:2015bma, Wang:2014hta}). However, among these solutions, only one is physically reasonable. Thus, finding all the possible solutions is the first step to properly decoding the experimental data.

The multiple solutions problem resulting from two resonances was discussed in Refs.~\cite{Zhu:2011ha,Bukin:2007kx}, wherein the authors indicated that there are two different solutions with equal fitting quality when fitting experimental data with two resonances, and they provided some analytical formulas to derive another one from a known solution. In Ref.~\cite{Han:2018wbo}, they extended to the case of three resonances and obtained some constraint equations between four different solutions. In the present work, we further extend the investigation to the case of an arbitrary number of resonances. By analyzing the mathematical structure of the amplitude, we present the source of multiple solutions for interfering resonances and obtain some simple relations between different solutions for more than two resonances. With these relations, we can easily derive all other solutions from a known one. 

Aside from the resonance contributions, a background or a nonresonance contribution is also usually introduced when fitting the cross sections or invariant-mass distributions. This background can be separated into noninterfering and interfering backgrounds. The former does not interfere with the resonance contributions; thus, it will not change the shape of the Breit-Wigner distribution. For example, in Ref.~\cite{Wang:2014hta}, the background of $e^+ e^- \to \pi^+ \pi^- \psi(2S)$ resulting from the sideband background of $J/\psi$ and $\psi(2S)$ signals in the lepton pair invariant-mass spectrum is considered to be a noninterfering background. Because the noninterfering background does not interfere with the resonances, such a background will not produce multiple fitting solutions.

As for the interfering background, it interferes with resonance contributions; thus, it can change the line shape of the Breit-Wigner distribution and lead to multiple solutions. For example, in Ref. \cite{Chang-Zheng:2014haa}, the line shape of the cross sections for $e^+ e^- \to \pi^+ \pi^- h_c$ was reproduced as a coherent sum of one resonance near 4.2 \GeV ~and a three-body phase space as an interfering background, two different solutions are obtained~\cite{Chang-Zheng:2014haa}. Of course, it might be possible that both noninterfering and interfering backgrounds exist for a given process. For example, when investigating the $\mu^+ \mu^-$ invariant-mass spectra of the $B\to K \mu^+ \mu^-$ process \cite{Aaij:2013pta}, the background is parameterized by an Argus function \cite{Albrecht:1990am}. The $B$ meson decays into $K \mu^+ \mu^-$ via the weak interaction; thus, the background is divided into axial-vector and vector parts, where the former cannot interfere with the resonance in the $\mu^+ \mu^-$ invariant-mass spectra, while interference between the vector background and resonances should not be ignored because all resonances in the $\mu^+ \mu^-$ invariant mass spectrum are vector states. As discussed above, the noninterfering background will not generate multiple solutions when fitting the experimental data, thus we will ignore such a background and only discuss interfering backgrounds. For the interfering background, we try to reveal a criterion for multiple solutions and simple methods for deriving a new background with equal fitting quality.

This paper is organized as follows. After the Introduction, we take the case of two resonances as an example to clarify the source of multisolutions in Sec.~\ref{sec:two-BW}, and then extend our analysis to the case of more than two resonances in Sec. \ref{sec:Multi-BW}. In Sec. \ref{sec:BGR}, we discuss the interference between an interfering background  and resonances. A short summary is provided in Section \ref{sec:Summary}.

\section{Interference between two resonances \label{sec:two-BW}}

In particle physics, a resonance is usually described by a relativistic Breit-Wigner function, e.g.,
 \begin{eqnarray}
 F_{BW}(s,M,\Gamma) = \dfrac{1}{s-M^2+iM\Gamma}=\dfrac{1}{s-p},
 \end{eqnarray}
where $s$ is the square of the center-of-mass energy, and $M$ and $\Gamma$ are the mass and width of the resonance, respectively. $p=M^2-iM\Gamma$ is the pole of the Breit-Wigner function. It is interesting to note that the Breit-Wigner function is a circle with center $O^\prime (0,-{i}/{(2M\Gamma)})$ and radius ${1}/{(2M\Gamma)}$ in the complex plane when we extend $s$ from positive to negative infinity.  

In the present work, we take the resonance in the $e^+e^-$ annihilation process as an example, where the scattering amplitude is related to the Breit-Wigner function by
\begin{eqnarray}
\label{Eq:BW}
&&A(\sqrt{s},M,\Gamma,f_R,\phi) =\nonumber\\ &&\hspace{10mm}\dfrac{M}{\sqrt{s}}\sqrt{12\pi f_R \Gamma} F_{BW}(s,M,\Gamma) \sqrt{\dfrac{\Phi(\sqrt{s})}{\Phi(M)}} e^{i\phi},
\end{eqnarray}
where $f_R=\Gamma_{e^+e^-} \times \mathrm{BR}(R\to f)$, while $\Gamma_{e^+e^-}$ and $\mathrm{BR}(R\to f)$ are the dilepton width of the resonance and the branch ratio of $R \to f$, respectively. $\Phi(\sqrt{s})$ is the phase space of $e^+e^- \to f$ and $\phi$ is the phase angle, which is usually assumed to be independent on $s$. Taking the case of two resonances as an example, the cross section is
\begin{eqnarray}
\sigma(s) =\left | A(\sqrt{s},M_0,{\Gamma}_0,{f_R}_0,\phi_0) + A(\sqrt{s},M_1,{\Gamma}_1,{f_R}_1,\phi_1)  \right| ^2 .\nonumber\\\label{Eq:CS-2R}
\end{eqnarray}
As indicated in Ref. \cite{Zhu:2011ha}, one can find two different solutions with equal fitting quality. In this example, we can set $\phi_0=0$ in Eq. (\ref{Eq:CS-2R}) without loss of generality; thus, there are seven free parameters: $M_0$, $\Gamma_0$, ${f_R}_0$, $M_1$, $\Gamma_1$, ${f_R}_1$, and $\phi_1$. In the following, we will show how to infer another solution from a known solution. As indicated in Refs. \cite{Han:2018wbo, Zhu:2011ha} the masses and widths of the resonances involved in two sets of parameters are identical. Thus, we only need to determine how to obtain another set of parameters $f_R^\prime$ and $\phi_R^\prime$ from the known $f_R$ and $\phi_R$. These two sets of parameters satisfy
\begin{eqnarray}
\label{equ:condition_0}
&&\left| A(\sqrt{s},M_0,{\Gamma}_0,{f_R}_0,\phi_0) + A(\sqrt{s},M_1,{\Gamma}_1,{f_R}_1,\phi_1) \right| =   \nonumber\\
&&\hspace{0.5cm} \left|A(\sqrt{s},M_0,{\Gamma}_0,{f_R^\prime}_0,\phi'_0) + A(\sqrt{s},M_1,\Gamma_1,{f_R^\prime}_1,\phi'_1) \right|. \hspace{8mm}
\end{eqnarray}
Taking the amplitude expression defined in Eq.~(\ref{Eq:BW}) into the above identity and performing some simplification yields
\begin{eqnarray}\label{Eq:Con1}
\big| z_0 F_0(s) + z_1 F_1(s) \big| =  
\big| z'_0 F_0(s) + z'_1 F_1(s)\big|,
\end{eqnarray}
with
\begin{equation}\label{For:Coeff}
z_k = \sqrt{{\Gamma_k}{f_R}_{k}} \dfrac{M_k}{\sqrt{\Phi(M_k)}}e^{i\phi_k}.
\end{equation}
Hereafter, we use the notion $F_k(s)= F_{BW}(s,m_k,\Gamma_k)$ for simplicity.

Dividing by a factor $z_1 F_0(s)$ in both sides of Eq.~(\ref{Eq:Con1}) yields
\begin{eqnarray}
\label{Eq:Con2}
\left|\frac{z_0}{z_1} + \frac{F_1(s)}{F_0(s)}\right| =\left|\frac{z'_1}{z_1}\right| \left|\frac{z'_0}{z'_1} + \frac{F_1(s)}{F_0(s)}\right|.
\end{eqnarray}
One should notice that the ratio between two Breit-Wigner functions satisfies 
\begin{eqnarray}\label{for:bw_ratio}
\dfrac{F_1(s)}{F_0(s)} = 1-(p_0-p_1)F_1(s),
\end{eqnarray}
where $p_0$ and $p_1$ are the poles of $F_0(s)$ and $F_1(s)$, respectively. With the above identity, we
can further simplify Eq. (\ref{Eq:Con2}) as follows:
\begin{eqnarray}
\Big|P-F_1(s) \Big| = \Big|\dfrac{z'_1}{z_1}\Big| \Big|P'-F_1(s) \Big|, \label{Eq:Con3}
\end{eqnarray}
where
\begin{eqnarray}\label{for:inverse_point}
 P  &=& \dfrac{z_0/z_1+1}{p_0-p_1}, \nonumber\\
 P' &=& \dfrac{z'_0/z'_1+1}{p_0-p_1}.
\end{eqnarray}

\begin{figure*}[!htpb]
\centering
\includegraphics[width=\textwidth]{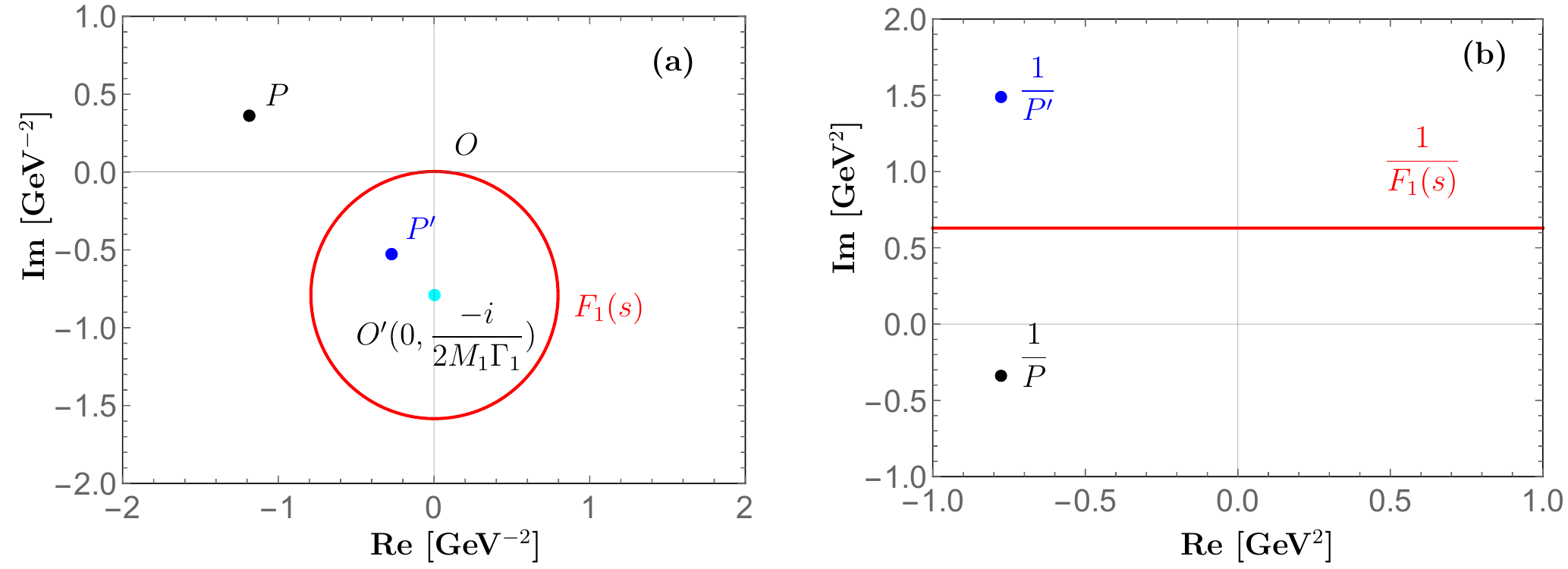}
\caption{Diagrams showing the relationship between $P$ and $P^\prime$ [diagram (a)] and their reciprocal [diagram (b)]. \label{Fig:Sketch}}
\end{figure*}

The geometric meaning of Eq. (\ref{Eq:Con3}) is that the ratio between the distances from any point on the circle defined by $F_1(s)$ to $P$ and $P'$ is a constant, which is $|{z'_1}/{z_1}|$. In other words, $P$ and $P'$ are a pair of reflection points on the circle, as shown in Fig. \ref{Fig:Sketch}-(a). The ratio of the distances in Eq.~(\ref{Eq:Con3}) can be evaluated as follows:
\begin{equation}\label{Eq:Distance_Ratio}
\Big|\dfrac{z'_1}{z_1}\Big|=\dfrac{|O P|}{|O P'|}.
\end{equation}
where $O$ is the origin of the complex plane and on the circle of $F_1(s)$. Furthermore, if $P$ and $P'$ are a pair of reflection points on the circle of $F_1(s)$, $1/P$ and $1/P^\prime$ are a pair of reflection points on $1/F_1(s)$, which is a line parallel to the real axes, as shown in Fig. \ref{Fig:Sketch}-(b). The reciprocals of $P$ and $P^{\prime}$  are simply connected by
\begin{eqnarray}
\label{Eq:Recip}   
\dfrac{1}{P^\prime}+p_1 = \dfrac{1}{P^*}+p^*_1,	
\end{eqnarray}
which will greatly simplify our estimation of $P^\prime$ from the known value of $P$.

\begin{figure*}[!htpb]
\centering
\includegraphics[width=\textwidth]{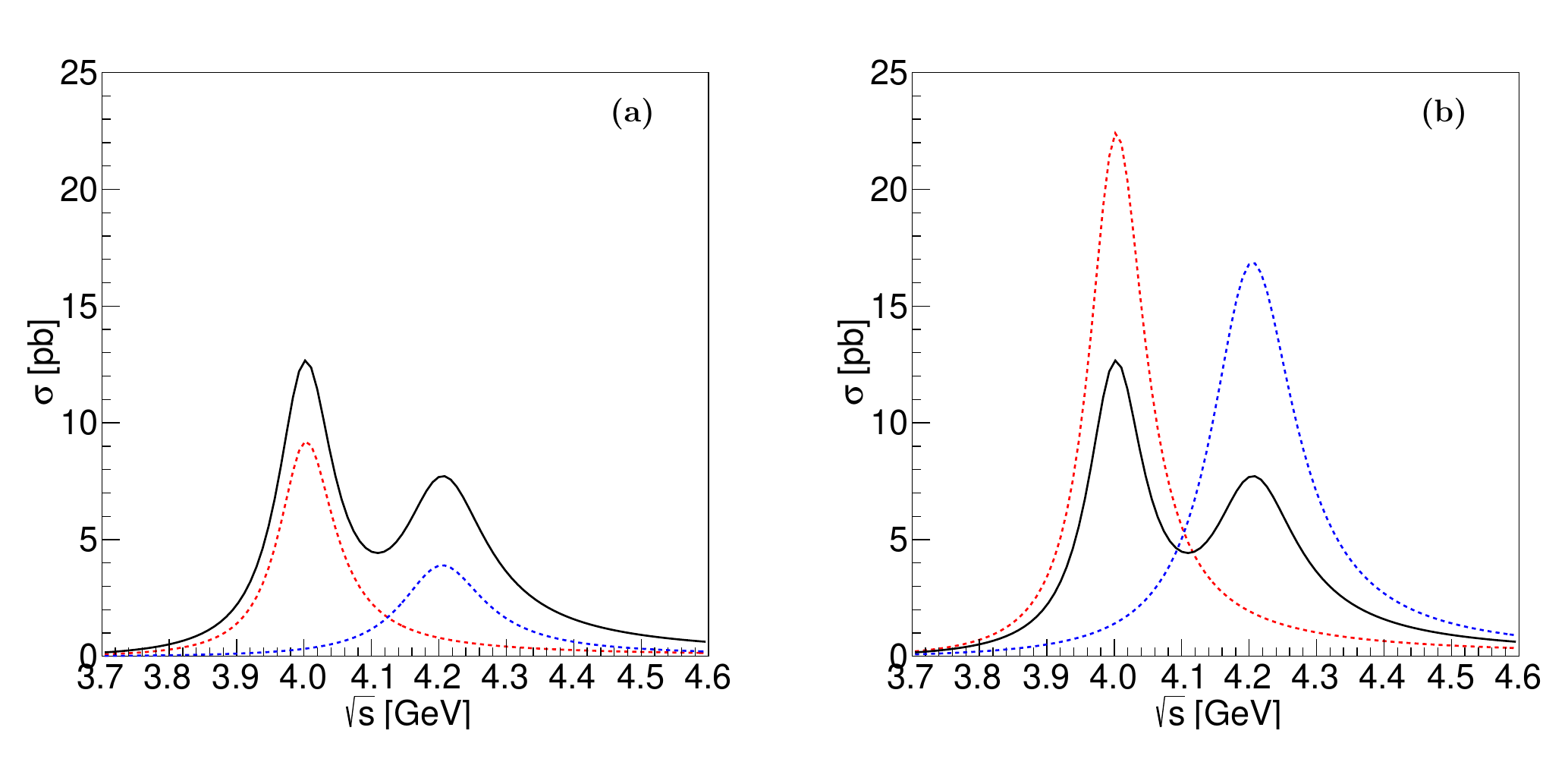}\caption{An example of interference between two Breit-Wigner distributions. The solid curve is the total cross section and the dashed curves are the individual contributions from the resonances.  \label{Fig:2BW}}
\end{figure*}

Hereafter, we take ~$e^+ e^- \to \pi^+ \pi^- J/\psi$ as an example to illustrate interference between two Breit-Wigner functions. We do not use the real experimental data here. Rather, we assume the masses and widths of two resonances to be $M_0=4000 \ \mathrm{\MeV},\ \Gamma_0=100 \ \mathrm{\MeV},\ M_1=4200\ \mathrm{\MeV}$ and $\Gamma_1=150 \mathrm{\MeV}$, respectively. We take $f_{R_0}=1.00 \ \mathrm{\eV}$, $f_{R_1}=0.700 \ \mathrm{\eV}$, and $\phi_1=\pi/4$ as inputs. The interference between these two states is constructive, as shown in Fig.\ref{Fig:2BW}-(a). In the following, we will show the process used to infer another set of parameters from these inputs.

\begin{itemize}[labelsep = .5em, leftmargin = 0pt, itemindent =1em]
	\item {\it \uline{ Recombination of input variables.}} --By using the definitions in Eqs. (\ref{For:Coeff}) and (\ref{for:inverse_point}), we can estimate the coordinates of $P$ and $O^\prime$, which are $P (-1.18,0.357)$ and $O^\prime(0,-0.794)$, respectively, in the present example. 
	\renewcommand{\ULdepth}{5pt}
\item {\it \uline{Estimating the coordinates of $P'$ and the values of $|{z'_1}/{z_1}|$ and ${z'_0}/{z'_1}$}. }  --The coordinates of $P'$ can be evaluated using Eq.~(\ref{Eq:Recip}), which is $(-0.274, -0.527)$. Then $|{z'_1}/{z_1}|$ can be estimated with Eq.~(\ref{Eq:Distance_Ratio}), yielding $|{z'_1}/{z_1}|=2.08$. With Eq.~(\ref{for:inverse_point}), one can obtain ${z'_0}/{z'_1}=-0.430+0.802~i$.
\item {\it \uline{Determination of $f'_{R_k}$ and $\phi_k^\prime$.}} --As for the new parameters, we can also set $\phi_0^\prime=0$ without loss of generality, as it can be a global phase with any value. Then, $z_0^\prime$ will be positive real. With the values of $|z_1^\prime/ z_1|$ and $z_0^\prime/z_1^\prime$, we find that $z_0^\prime=5.67\times 10^{4} ~\mathrm{\MeV^2}$ and $z_1^\prime=(-2.94 -5.49 i)\times 10^{4} ~\mathrm{\MeV^2}$, respectively. With Eq. (\ref{For:Coeff}), we can determine another set of parameters, which are $f_{R_0}^\prime=2.43\ \mathrm{\eV}$, $f_{R_1}^\prime=3.03\ \mathrm{\eV}$, and $\phi_1^\prime=4.22$. In this case, the interference between these two resonances is destructive, as shown in Fig.~\ref{Fig:2BW}-(b).
\end{itemize}

\section{Combination of more than two resonances \label{sec:Multi-BW}}
Now we extend our analysis in the above section to the case of more than two resonances. Here, we consider interference between $n+1, (n \ge 2)$ resonances. In this case, Eq. (\ref{Eq:Con1}) becomes
\begin{eqnarray}
\label{Eq:Multi-Con1}
\left|\sum^{n}_{k=0} z_k F_{k}(s) \right|= \left|\sum^{n}_{k=0} z'_k F_{k}(s) \right|.
\end{eqnarray}
Dividing by a factor $F_0(s)$ in both sides of Eq. (\ref{Eq:Multi-Con1}) yields
\begin{eqnarray}\label{Eq:Multi-Con2}
\left|z_0+\sum^{n}_{k=1} z_k \dfrac{F_k(s)}{F_0(s)} \right| = \left|z'_0+\sum^{n}_{k=1} z'_k \dfrac{F_k(s)}{F_0(s)} \right|.
\end{eqnarray}
Similar to Eq. (\ref{Eq:Con3}), the above equation can be further formalized as follows:
\begin{eqnarray}
\label{Eq:Multi-Con3}
\left|A_0+\sum^{n}_{k=1} A_k F_k(s)\right| = \left|A'_0+\sum^{n}_{k=1} A'_k F_k(s)\right|,
\end{eqnarray}
where $A_0 = \sum^n_{i=0} z_i$ and $A_k = z_k(p_k-p_0)$, respectively. As indicated in the Appendix \ref{Sec:App}, one can further factorize the left side of Eq. (\ref{Eq:Multi-Con3}) into the following form: 
\begin{eqnarray}
\label{Eq:Fac}
A_0+\sum^{n}_{k=1} A_k F_k(s)  = A_0\prod^n_{i=1}(1+a_iF_i(s)),
\end{eqnarray}
where $a_i$ can be estimated from $A_k$. Similar to the case of interference between two resonances, we should obtain all sets of fitting parameters or $a_i^\prime$ with equally fitting quality from the already obtained fitting parameters or $a_i$, i.e., 
\begin{eqnarray}
\left|A_0\prod^n_{i=1}(1+a_iF_i(s)) \right| =\left|A_0^\prime \prod^n_{i=1}(1+a_i^\prime F_i(s)) \right|.
\label{Eq:Multi-Con4}
\end{eqnarray}
One should notice that $F_i\to 0$ when $s\to \infty$; thus, we have $|A_0| =|A_0^\prime|$. Equation (\ref{Eq:Multi-Con4}) can then be further simplified to
\begin{eqnarray}
\left|  \prod^n_{i=1} F_i(s)(a_i+F^{-1}_i(s)) \right| =\left| \prod^n_{i=1}  F_i(s)(a'_i + F^{-1}_i(s)) \right|.
\label{Eq:Multi-Con5}
\end{eqnarray}
Comparing the factors in both sides with index~$i$, it is clear that $-a_i$ and $-a'_i$ are a pair of reflections at $F^{-1}_i(s)$; thus, $a'_i$ can be derived using Eq. (\ref{Eq:Recip}), yielding 
\begin{equation}\label{Eq:a_parameter}
-a'_i+p_i = -a^*_i+p^*_i.
\end{equation}

 \begin{figure*}[htb]
 \includegraphics[width=\textwidth]{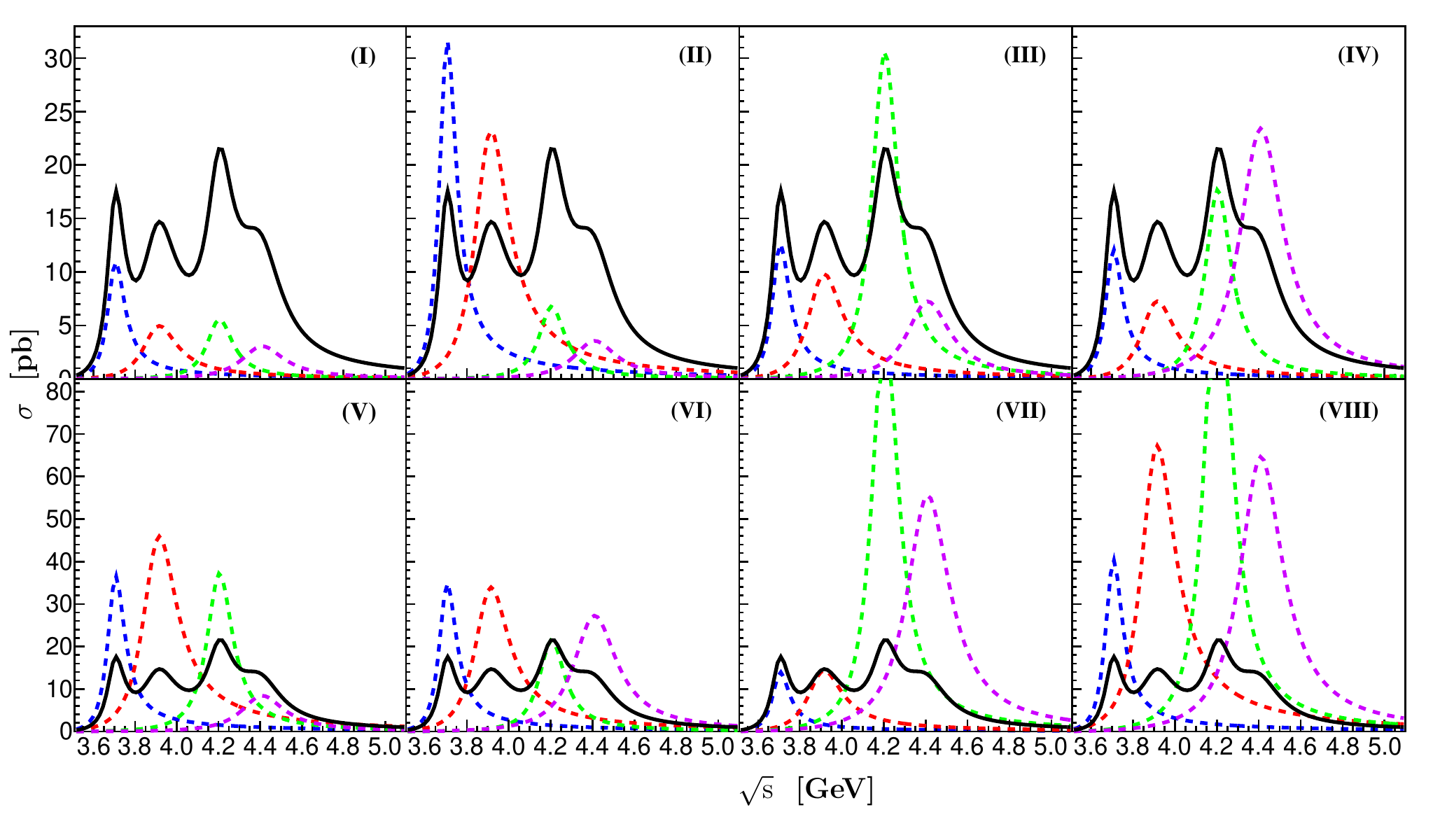}
 \caption{Cross sections with interference between four resonances. Diagram (a) corresponds to input parameters, and diagrams (b)--(h) denote solutions II--VIII estimated with the present methods. \label{Fig:4R} }
 \end{figure*}

\begin{table*}[htb]
\caption{Parameters estimated with the present method from one set of input parameters (Solution I). \label{Tab:Para}}

\begin{tabular}{p{2.2cm}<{\centering} p{1.6cm}<{\centering} p{1.6cm}<{\centering} p{1.6cm}<{\centering} p{1.6cm}<{\centering} p{1.6cm}<{\centering} p{1.6cm}<{\centering} p{1.6cm}<{\centering} p{1.6cm}<{\centering}}
\toprule[1pt]\toprule[1pt]
 Solutions  & I (Input) & II & III & IV & V & VI & VII & VIII\\
 \midrule[1pt]
$M_0$ (\MeV)  & \multicolumn{8}{c}{3700}  \\
$\Gamma_0$ (\MeV) & \multicolumn{8}{c}{100}  \\
$\phi_0 $ (rad) &  \multicolumn{8}{c}{0}  \\  
$f_{R_0}$ (\eV) & 1.00  & 2.89 & 1.15  & 1.10 & 3.33 & 3.19 & 1.27 & 3.68  \\
\midrule[1pt]  
$M_1$ (\MeV) &  \multicolumn{8}{c}{3900}\\
$\Gamma_1$ (\MeV) &\multicolumn{8}{c}{200}\\
$\phi_1 $ (rad)& 0.70 & 3.86& 1.08& 0.948& 4.24 & 4.11 & 1.33 & 4.49\\  
$f_{R_1}$ (\eV)& 1.00 & 4.68 & 1.98 & 1.46& 9.27 & 6.85& 2.90 & 13.6\\
\midrule[1pt]  
$M_2$ (\MeV) & \multicolumn{8}{c}{4200}\\
$\Gamma_2$ (\MeV) & \multicolumn{8}{c}{150}\\
$\phi_2 $ (rad)& 1.40 & 5.30 & 4.43 & 2.05 & 2.84 & 0.462 & 5.87 & 3.49\\  
$f_{R_2}$ (\eV)& 1.00&1.21 & 5.50 & 3.18& 6.68& 3.86 &  17.5 & 21.2\\
\midrule[1pt]  
$M_3$ (\MeV) & \multicolumn{8}{c}{4400}\\
$\Gamma_3$ (\MeV) &\multicolumn{8}{c}{250}\\
$\phi_3 $ (rad)& 2.1 & 6.19& 0.310& 5.74 & 4.40 & 3.54& 3.95& 1.75\\  
$f_{R_3}$  (\eV)& 1.0 & 1.16& 2.37& 7.67& 2.76 & 8.93& 18.2& 21.2\\
\bottomrule[1pt] \bottomrule[1pt]
\end{tabular}

\end{table*}
 
For the of case of $n+1$ resonances, one can determine a new set of fitting parameters by replacing any subset of $\{a_i\}$ by their reflections $\{a'_i\}$. There are $\sum_{i=1}^n C_n^i$ such replacements, and we finally have $1+\sum_{i=1}^n C_n^i=2^n$ sets of parameters with equal fitting quality.

We generate the cross sections with a combination of four resonances. Taking the parameters of solution I in Table \ref{Tab:Para} as inputs, we can estimate the remaining seven sets of parameters, which are also listed in Table \ref{Tab:Para} as solutions II-VIII. The cross sections and individual resonance contributions are presented in Fig. \ref{Fig:4R}, where diagrams (a)--(h)  correspond to solutions I--VIII, respectively. From the figure, one can find the resonance contributions are rather different in different solutions.

\section{Interference with background \label{sec:BGR}}

We analyzed interference between resonances in the preceding sections. In the present section, we further discuss interference between the background and resonances. Such kind of interference is much more complicated than interference between resonances because the physical origin of the background is usually unclear and the mathematical formalization is not well determined. Some simple functions, such as polynomial and exponential functions with real parameters, are often used to describe the background.

Here, we consider an amplitude function $D(s)$, which is a combination of Breit-Wigner functions $F_k(s)$ and an interfering background function $B(s)$, i.e.,
\begin{equation}
D(s)=z_0 B(s)+\sum^{k=n}_{k=1}z_k F_k(s).
\end{equation}
Here the Breit-Wigner functions' coefficients $z_k$ are defined as the same as that in Eq.~(\ref{For:Coeff}). The cross section and $D(s)$ are connected by
\begin{eqnarray}\label{Eq:XS_DS}
\sigma=\dfrac{12\pi \Phi(\sqrt{s})}{s}|D(s)|^2.
\end{eqnarray}

The background function here usually has no essential singularity in any finite domain, and the Breit-Wigner function $F_k$ has a pole $p_k$. Assuming $D_s$ has zeros at $\{q_1, q_2, \dots \}$, with the help of the Weierstrass factorization theorem \cite{Weierstrass}, the amplitudes can be factorized as 
\begin{eqnarray}\label{Eq:FACT_I}
D(s)= \Big[\prod _i (s-q_i) \Big] H(s).
\end{eqnarray}
Thus, if one replaces any $q_i$ by its complex conjugate, i.e., $q_i^\ast$, the modulus of $D(s)$ will not change if $s$ is real. Here, we take one of the zeros $q_m$ as an example to illustrate such a replacement. Because $q_m$ is the zero point of $D(s)$, we have
\begin{equation}
D(q_m)=z_0 B(q_m)+\sum^{k=n}_{k=1}z_k F_k(q_m)=0,
\end{equation}   
and $D(s)$ can be factorized as follows:
\begin{eqnarray} 
D(s)=(s-q_m) G(s), \label{Eq:BKG-Fac}
\end{eqnarray}
where $G(s)= \big[\prod_{i,i\neq m} (s-q_i) \big] H(s)$. One should notice that~$G(s)$ might still have zeros other than $q_m$, while $H(s)$ has no zeros.\par
In principle, with the Weierstrass factorization theorem, one can determine the particular expression of $G(s)$ for a given $B(s)$. Here, we notice that the Breit-Wigner function satisfies
\begin{equation}
     F_k(s)-F_k(q_m) = -\dfrac{s-q_m}{(s-p_k)(q_m-p_k)},
\end{equation}
then, we have
\begin{eqnarray}
F_k(s)  &=& -\frac{s-q_m}{q_m -p_k} F_k(s) +\frac{1}{q_m -p_k}\nonumber\\
                        &=&-\frac{s-q_m}{q_m -p_k} F_k(s) +F_k(q_m). 
\label{Eq:BGK-Fk}
\end{eqnarray}
The above property of Breit-Wigner functions yields
\begin{eqnarray}
\label{eqn:H_function}
&& G(s)= \dfrac{D(s)}{s-q_m}\nonumber\\
&&     = \dfrac{z_0}{s-q_m}B(s) +\dfrac{1}{s-q_m}\sum^{k=n}_{k=1}z_k F_k(q_m)-\sum^{k=n}_{k=1}\frac{z_k}{(q_m-p_k)} F_k(s) \nonumber\\ 
&& = \dfrac{z_0}{s-q_m}  \Big[ B(s) -B(q_m)  \Big]- \sum^{k=n}_{k=1}\frac{z_k}{(q_m-p_k)} F_k(s).
 \end{eqnarray}
One can find the structure of $G(s)$ is the same as that of $D(s)$ if one treats the first term as an interfering background function. Replacing $(s-q_m)$ by $(s-q_m^\ast)$ in Eq. (\ref{Eq:BKG-Fac}), we can obtain a new amplitude $D^\prime (s)$: 
\begin{eqnarray}
&&D^\prime(s)=(s-q_m^\ast)G(s)  \nonumber \\
&&= \dfrac{z_0(s-q_m^\ast)}{s-q_m}\Big[B(s)-B(q_m)\Big]-\sum^{k=n}_{k=1}\frac{z_k(s-q_m^\ast)}{(q_m-p_k)} F_k(s)\nonumber\\
&&=z_0 B(s)+\dfrac{z_0(B(s)-B(q_m))(q_m-q^*_m)}{s-q_m} \nonumber\\
&&+ \sum^{k=n}_{k=1}\dfrac{z_k(q_m^*-p_k)}{q_m-p_k}F_k(s).
\end{eqnarray}
This new amplitude has the same modulus as $D(s)$, so $D^\prime(s)$ has the same cross section as $D(s)$. The above analysis can be summarized as transformations on $z_k$ and $B(s)$,
\begin{equation}  
\left\{  
\begin{array}{ccl}
B(s)&\rightarrow & B(s)^\prime = B(s) + \dfrac{(q_m-q^*_m)[B(s)-B(q_m)]}{s-q_m}, \\
z_k &\rightarrow & z_k^\prime = \dfrac{q^*_m-p_k}{q_m-p_k}z_k. 
\end{array}
\right.  
\label{Eq:Tran}
\end{equation}
This transformation can be applied for each zero. If there are $n$ zeros in $D(s)$, there apparently are $2^n$ combinations with the same expected cross section because each zero doubles the multiplicity of combinations \footnote{ 
The transformation in Eq.~(\ref{Eq:Tran}) can also be applied to the case of interference between multiple Breit-Wigner functions, as discussed in Sec.~\ref{sec:Multi-BW}. On both sides of Eq. (~\ref{Eq:Multi-Con3}), $A_0$ and $A'_0$ can be treated as backgrounds. The transformation on $A_0$ and $A_k$, presented in Eq. (\ref{Eq:Trans_A}) is consistent with the transformation on Eq. (\ref{Eq:Tran}).}.

Here, we present examples with two kinds of backgrounds to illustrate the transformations discussed above, which are exponential and polynomial forms of the interfering background. In the first case, the amplitude $D(s)$ is in the form
\begin{eqnarray}
D(s)=z_0e^{-s/\Lambda^2} + z_1 \dfrac{1}{s-p} \label{Eq:BGK-1}.
\end{eqnarray}

\begin{figure*}[!htpb]
     \includegraphics[width=1\textwidth]{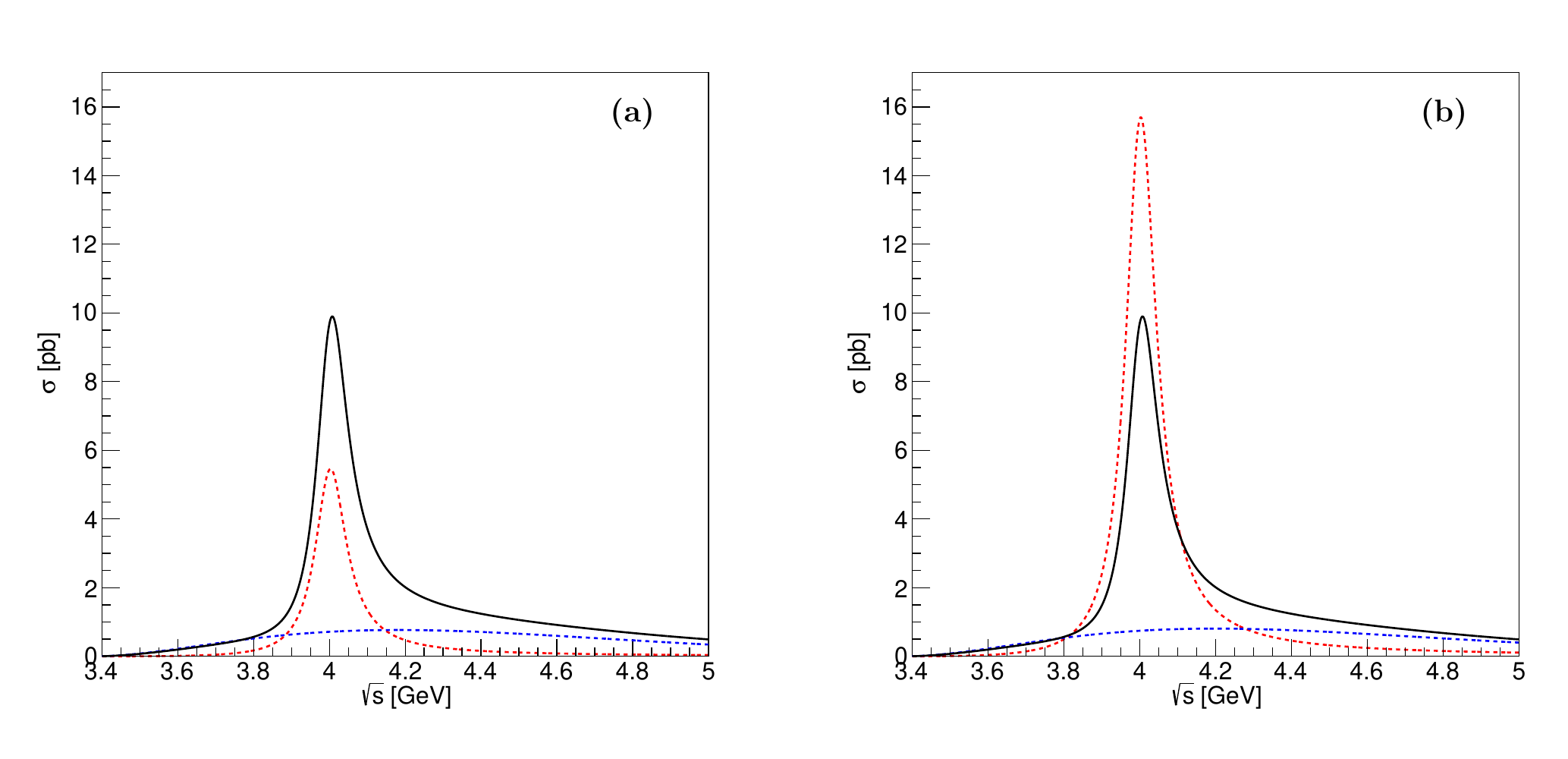}
\caption{Interference between a resonance and an exponential background. The dashed and solid curves show the individual contributions and the cross sections obtained with the input parameters (diagram (a)) and the new amplitudes obtained with the present method (diagram (b)). \label{fig:exp_bkg}}
\end{figure*}

The input parameters are $m_R=4.0 \ \mathrm{\GeV}$, $\Gamma_R =0.1 \ \mathrm{\GeV} $, $f_{R}=0.592 \mathrm{\eV}$, $\phi_R=1.11 \ \mathrm{rad} $,  $\Lambda =3\ \mathrm{\GeV}$, and $z_0 = 0.15$, respectively. One can estimate $z_1 = (0.0125+0.025~i) \ \mathrm{\GeV^2} $ with Eq. (\ref{For:Coeff}). One zero of $D(s)$ is found at $q =(15.41-1.25~i) \ \mathrm{\GeV}^2$. One can calculate a new amplitude from the replacements in Eq. (\ref{Eq:Tran}):
\begin{eqnarray}
D^\prime(s) = \left[z_0e^{-s/\Lambda^2} + \dfrac{z_0(q-q^*)(e^{-s/\Lambda^2}-e^{-q/\Lambda^2})}{s-q}\right]+\dfrac{z'_1}{s-p} ,\nonumber\\
\label{Eq:BGK-1}
\end{eqnarray}
where $z'_1 = (0.0219-0.0420~i) \ \mathrm{\GeV}^2$. The corresponding new resonance parameters change to $f'_R = 1.70$ eV and $\phi' = 5.19$ rad. As a comparison, we present the contributions and cross sections in Fig. \ref{fig:exp_bkg}. One can find that the background contributions are very similar for both amplitudes. However, the phase angle of these two set solutions are not equal, which indicates interferences between the background and different resonances are much different, leading to very different resonance contributions. In this example, the explicit expression for the interfering background is changed.

\par

\begin{figure*}[htb]
 \includegraphics[width=\textwidth]{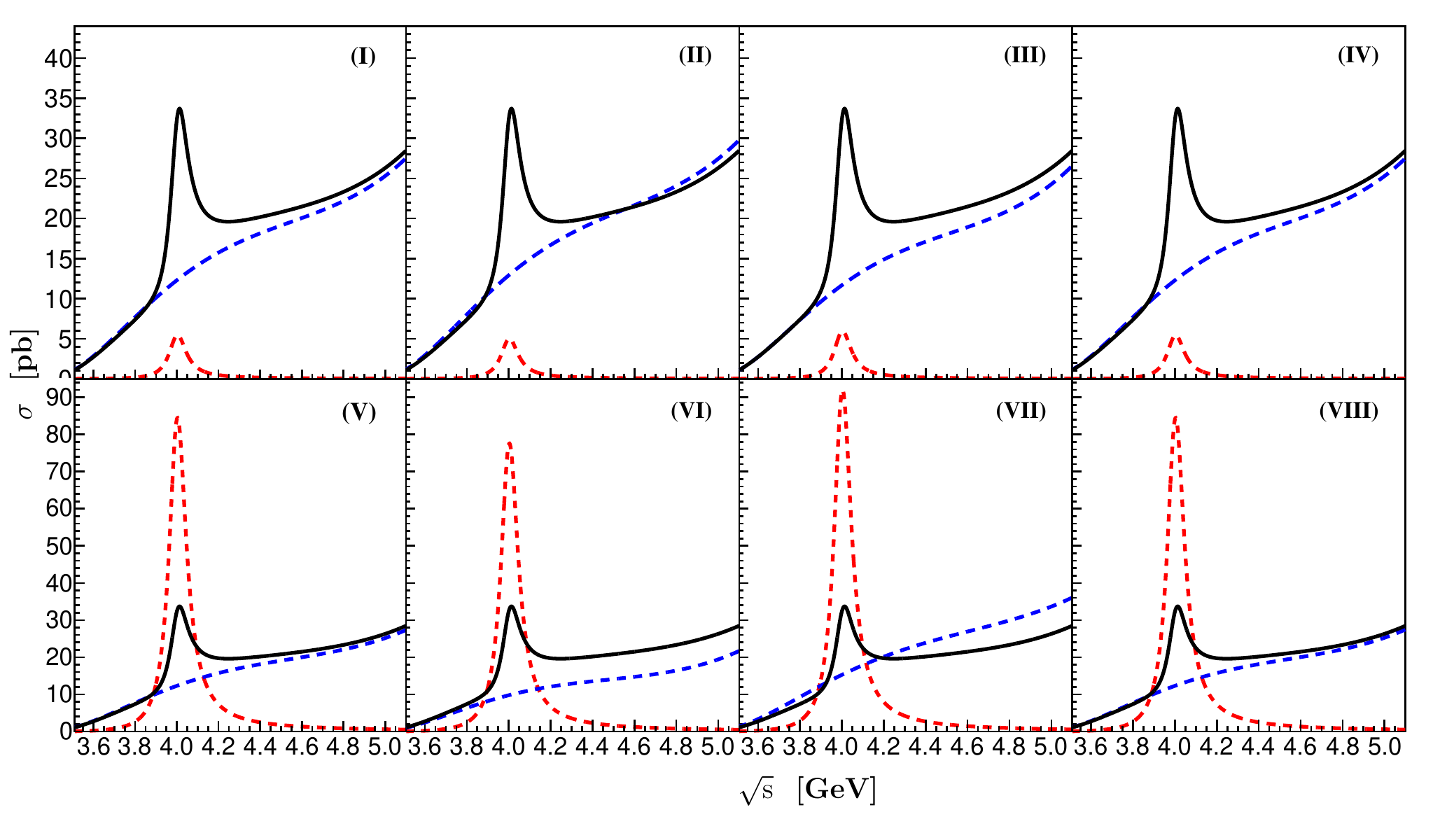}
 \caption{The same as Fig. \ref{fig:exp_bkg} but for polynomial background. \label{Fig:PolyBKG} }
 \end{figure*}

In the second example, the amplitude $D(s)$ is the coherent sum of a second order polynomial and one Breit-Wigner function:
\begin{eqnarray}
D(s)= w_2s^2+w_1s+w_0 + z_1 \dfrac{1}{s-p}. \label{Eq:BGK-2}
\end{eqnarray}
With the input parameters (solution I) in Table \ref{Tab:PolyBkg}, one finds that there are three zeros in the amplitude, which are $q_1 = (15.87-0.63~i) \ \mathrm{\GeV}^2$, $q_2 =( 23.99+14.25~i) \ \mathrm{\GeV}^2$, and $q_3 = (24.13-14.02~i) \ \mathrm{\GeV}^2$. One can use the method discussed here to determine seven other sets of parameters with moduli identical to that already known. In the present example, after performing the transformation in Eq. (\ref{Eq:Tran}), one finds that the new background is still a second degree polynomial, but with complex coefficients. We summarize all the parameters in Table \ref{Tab:PolyBkg}. The cross sections and individual contributions are presented in Fig.~\ref{Fig:PolyBKG}. One can find that the parameter $f_R$ changes dramatically in solutions V, VI, VII, and VIII. Such a phenomenon results from the fact that the zero $q_1$ is very close to the pole of the Breit-Wigner function, which has a significant effect in the transformation, as shown in Eq.~(\ref{Eq:Tran}). 

\begin{table*}[htb]
\caption{Parameters estimated with the present method from one set of input parameters (Solution I) with a 2nd degree polynomial background. \label{Tab:PolyBkg}}

\begin{tabular}{p{2cm}<{\centering} p{1.4cm}<{\centering} p{1.4cm}<{\centering} p{1.4cm}<{\centering} p{1.4cm}<{\centering} p{1.4cm}<{\centering} p{2.4cm}<{\centering} p{2.4cm}<{\centering}}
\toprule[1pt]\toprule[1pt]
\multirow{2}{*}{ Solutions}  & $m_1$& $\Gamma_1$  & $\phi_1$ & $f_{R_1}$  & $10^4\times w_2$ & $10^2\times w_1$ & \multirow{2}{*}{$10\times w_0$}\\
 &  (\MeV) & (\MeV) &  (rad) & (\eV) & ($\GeV^{-4}$) & ($\GeV^{-2}$) & \\
\midrule[1pt]  
I &\multirow{8}{*}{4000} & \multirow{8}{*}{100} & $1.11$ & $0.592$ & \multirow{8}{*}{ $4.00$ } & $-1.92$  & $3.10$ \\
II &  & & 5.27 &0.543 && $-1.92-1.14i$  & $1.47 -2.74i$\\
III & &  & 3.20 & 0.645 &   &$-1.92-1.12i$ & $1.53+2.68i$ \\
IV &  &  & 1.08 & 0.592 &   &$-1.92+0.02i$ & $3.10-0.60i$ \\
V  &  &  & 4.87 & 9.18 &   &$-1.92-0.05i$ & $3.10+1.62i$ \\
VI &  &  & 2.76 & 8.43 &   &$-1.92+1.09i$ & $1.62-2.58i$ \\
VII&  &  & 0.68 & 10.00 &   &$-1.92-1.17i$ & $1.38+2.84i$ \\
VIII& &  & 4.85 & 9.18 &   &$-1.92-0.03i$ & $3.10+1.02i$\\
\bottomrule[1pt] \bottomrule[1pt]

\end{tabular}
\end{table*}
\par

 \section{Conclusions\label{sec:Summary}}
 
Different solutions with equal fitting quality can be obtained when one fits the experimental data with the coherent sum of several Breit-Wigner functions or interference between a background and resonances. We first take two resonances as an example to analyze the source of multiple solutions. By analyzing the mathematical structure of the amplitudes, we find that the Breit-Wigner function $F_{k}(s)$ is a circle in the complex plane, which indicates that there should exist a pair of reflection points $(P, P^\prime)$ in the complex plane, while the ratio between the distances from these two points to any point on the circle is a constant. Furthermore, one can find $(1/P,1/P^\prime)$ is also a pair of reflection points at $1/F(s)$, while $1/F(s)$ is a line parallel to the real axis, which will further simplify our estimation.

In a similar way, we extend such an analysis to the case of more than two resonances. We find that the amplitude can be factorized after some mathematical simplification. Furthermore, every factor is a constant plus a Breit-Wigner function. We can then extend the methodology used to analyze the case of two resonances to one involving more than two resonances. We can easily infer all other solutions from the known solution with equal fitting quality using the present method.
 
We also discussed interference between background and resonances. Generally, the source of an interfering background is very complicated. With the help of the Weierstrass factorization theorem, we can factorize the amplitude, which is the coherent sum of Breit-Wigner functions and an interfering background. Assuming the amplitude has a zero $q$, then, one can obtain a new amplitude with the same modulus as the old amplitude through a simple replacement, i.e., $(s-q)\to (s-q^\ast)$. We can determine a new background and new parameters with equal fitting quality. 

The programs related to examples in the present work are available on GitHub \cite{website}.

 \section*{Acknowledgement}
 We would like to thank Xiao-Rui Lu and Cheng-Ping Shen for useful discussions. This work is supported in part by the National Natural Science Foundation of China (NSFC) under Grants No. 11705028, 11775050, and 11375240, and by the Fundamental Research Funds for the Central Universities.


\appendix

\section{Coefficients $a_i$ \label{Sec:App}}

In this Appendix, we present a procedure for determining the coefficients $a_i$ in Eq. (\ref{Eq:Fac}). Using the property of Breit-Wigner functions $F_i F_j = (F_i-F_j)/(p_i-p_j)$ and comparing the coefficients on both sides of Eq. (\ref{Eq:Fac}) yields the following equations:
\begin{eqnarray}  
\left\{  
\begin{array}{rcl}
a_1 (\dfrac{a_2}{p_1-p_2}+1)(\dfrac{a_3}{p_1-p_3}+1)\cdots(\dfrac{a_n}{p_1-p_n}+1) &=& \dfrac{A_1}{A_0} ,\\
(\dfrac{a_1}{p_2-p_1}+1)a_2(\dfrac{a_3}{p_2-p_3}+1)\cdots(\dfrac{a_n}{p_2-p_n}+1) &=& \dfrac{A_2}{A_0},\\
\vdots  \\
(\dfrac{a_1}{p_n-p_1}+1)(\dfrac{a_2}{p_n-p_2}+1)(\dfrac{a_3}{p_n-p_3}+1)\cdots a_n &=& \dfrac{A_n}{A_0}.
\end{array}
\right.  \nonumber\\
\label{eqn:coefficients_1}
\end{eqnarray}
For simplify, we further define $D_k = A_k/A_0 \prod^n_{i=1,i\neq k}(p_k-p_i)$ and $\alpha_k = a_k - p_k$. Substituting these identities  into Eq. (\ref{eqn:coefficients_1}) yields:
\begin{eqnarray}  
\left\{  
\begin{array}{rcl}
\prod^n_{i=1}(\alpha_i+p_1) &=& D_1, \\
\prod^n_{i=1}(\alpha_i+p_2) &=& D_2 ,\\
   \vdots \\
   \prod^n_{i=1}(\alpha_i+p_n) &=& D_n.
\end{array}
\right.  
\label{eqn:coefficients_2}
\end{eqnarray}
Expanding the left side of the identities in the above equation yields:
\begin{equation}
\renewcommand*{\arraystretch}{1.5}
\label{Eq:Matrix}
\begin{pmatrix}
    1       & p_1 & p^2_1 & \dots & p^{n-1}_1 \\
    1       & p_2 & p^2_2 & \dots & p^{n-1}_2 \\
    \hdotsfor{5} \\
    1       & p_n & p^2_n & \dots & p^{n-1}_{n}
\end{pmatrix}
\begin{pmatrix}
X_{n}\\
X_{n-1}\\
\hdotsfor{1}\\
X_1
\end{pmatrix}
+
\begin{pmatrix}
p^n_1\\
p^n_2\\
\hdotsfor{1}\\
p^n_n
\end{pmatrix}
=
\begin{pmatrix}
D_1\\
D_2\\
\hdotsfor{1}\\
D_n
\end{pmatrix},
\end{equation}
where $X_k =\sum \prod_{i=1}^{k} \beta_{i}$ with $\{\beta_1,~\beta_2,~\cdots \beta_k\}$ is a $k$-element subset of  $\{\alpha_1, ~\alpha_2,~\cdots,~\alpha_n\}$. All equations in the above array are linear equations of $X_i$; thus, this array could be solved very easily.

From the definitions of $X_i$ and the Vieta theorem \cite{Viate}, one can find that $-\alpha_i,\ \{i=1,2,\cdots n\}$ are the roots of the following $n$th degree equation with one unknown:
\begin{eqnarray}
\label{eqn:characteristic_equation}
t^n+X_1 t^{n-1} + X_2 t^{n-2}\cdots X_{n-1} t + X_n = 0.
\end{eqnarray}
By solving the linear equation in Eq.~ (\ref{Eq:Matrix}) and the algebraic equation in Eq.~ (\ref{eqn:characteristic_equation}), one can determine $\alpha_i$ and $a_i$. One can then determine new sets of parameters $a'_i$ and $\alpha'_i$ using Eq.~(\ref{Eq:a_parameter}). It should be noticed that replacing $a_i$ with $a'_i$  is equivalent to replacing $\alpha_i$ with $\alpha^*_i$.
It is obvious that this replacement changes $D_m$ to $D_m^\prime$ in Eq. (\ref{eqn:coefficients_2}), where $D_m^\prime$ is
\begin{eqnarray}\label{Eq:Trans_D}
   D'_m = \dfrac{\alpha^*_{i}+p_m}{\alpha_{i}+p_m}D_m.
\end{eqnarray}
\par
According to the definition of $D_m$, one finds that $A_m/A_0$ also transforms to $A_m^\prime/A_0^\prime$, where $A_m^\prime/A_0^\prime$ is
\begin{eqnarray}\label{Eq:Trans_A}
   \dfrac{A'_m}{A'_0} = \dfrac{\alpha^*_{i}+p_m}{\alpha_{i}+p_m}\dfrac{A_m}{A_0}.
\end{eqnarray}
As we discussed in Sec. \ref{sec:Multi-BW}, one has $|A_0|=|A'_0|$. Here, we can set $A_0=A_0^\prime$ because the phases of $A_0$ and $A_0^\prime$ could be treated as global phases of the amplitudes. One can then determine the transformation $A_m \to A_m^\prime$, which is the same as replacing $z_k$ in Eq. (\ref{Eq:Tran}).
\par

\end{document}